\documentclass[3p,times,procedia]{elsarticle}
\usepackage{nupha_ecrc}
\usepackage[utf8]{inputenc}

\volume{00}

\firstpage{1}

\journalname{Nuclear Physics A}

\runauth{}

\jid{nupha}

\jnltitlelogo{Nuclear Physics A}

\usepackage{amssymb}

\usepackage[figuresright]{rotating}

\begin{document}

\begin{frontmatter}



\dochead{XXVIIIth International Conference on Ultrarelativistic Nucleus-Nucleus Collisions\\ (Quark Matter 2019)}

\title{NA61/SHINE results on fluctuations and correlations at CERN SPS energies}


\author{Maja Ma\'{c}kowiak-Paw{\l}owska\\ for the NA61/SHINE Collaboration}

\address{Faculty of Physics, Warsaw University of Technology, Koszykowa 75, 00-662 Warsaw}

\begin{abstract}
The aim of the NA61/SHINE strong interaction programme is to explore the phase diagram of strongly interacting matter. The main physics goals are the study of the onset of deconfinement and the search for the critical point of strongly interacting matter. These goals are pursued by performing a beam momentum (13$A$ -- 150/158$A$ GeV/$c$) and system size (p+p, p+Pb, Be+Be, Ar+Sc, Xe+La, Pb+Pb) scan. This contribution presents new results from NA61/SHINE on fluctuations and correlations which include in particular quantum correlations, as well as multiplicity and net-charge fluctuations, proton density fluctuations and anisotropic collective flow. Obtained results are compared with other experiments and with model predictions.
\end{abstract}

\begin{keyword}
critical point \sep onset of deconfinement \sep fluctuations \sep intermitency \sep femtoscopy \sep flow

\end{keyword}

\end{frontmatter}


\section{Introduction}
\label{sec:intro}
NA61/SHINE~\cite{Antoniou:2006mh,Abgrall:2014xwa} at the CERN Super Proton Synchrotron (SPS) is a fixed-target experiment pursuing a rich physics program including measurements for strong interactions, neutrino, and cosmic ray physics.

The strong interactions program focuses on search for the critical point (CP) and study of the onset of deconfinement (OD) of strongly interacting matter.  NA61/SHINE is the first experiment to perform a two-dimensional scan, in beam momentum (13$A$ -- 150/158$A$ GeV/$c$)  and  system size  (p+p, p+Pb, Be+Be, Ar+Sc, Xe+La, Pb+Pb) of colliding nuclei. 

\section{Search for the critical point}
\label{sec:CP}
\begin{figure}[ht]
\centering
\includegraphics[width=0.45\textwidth]{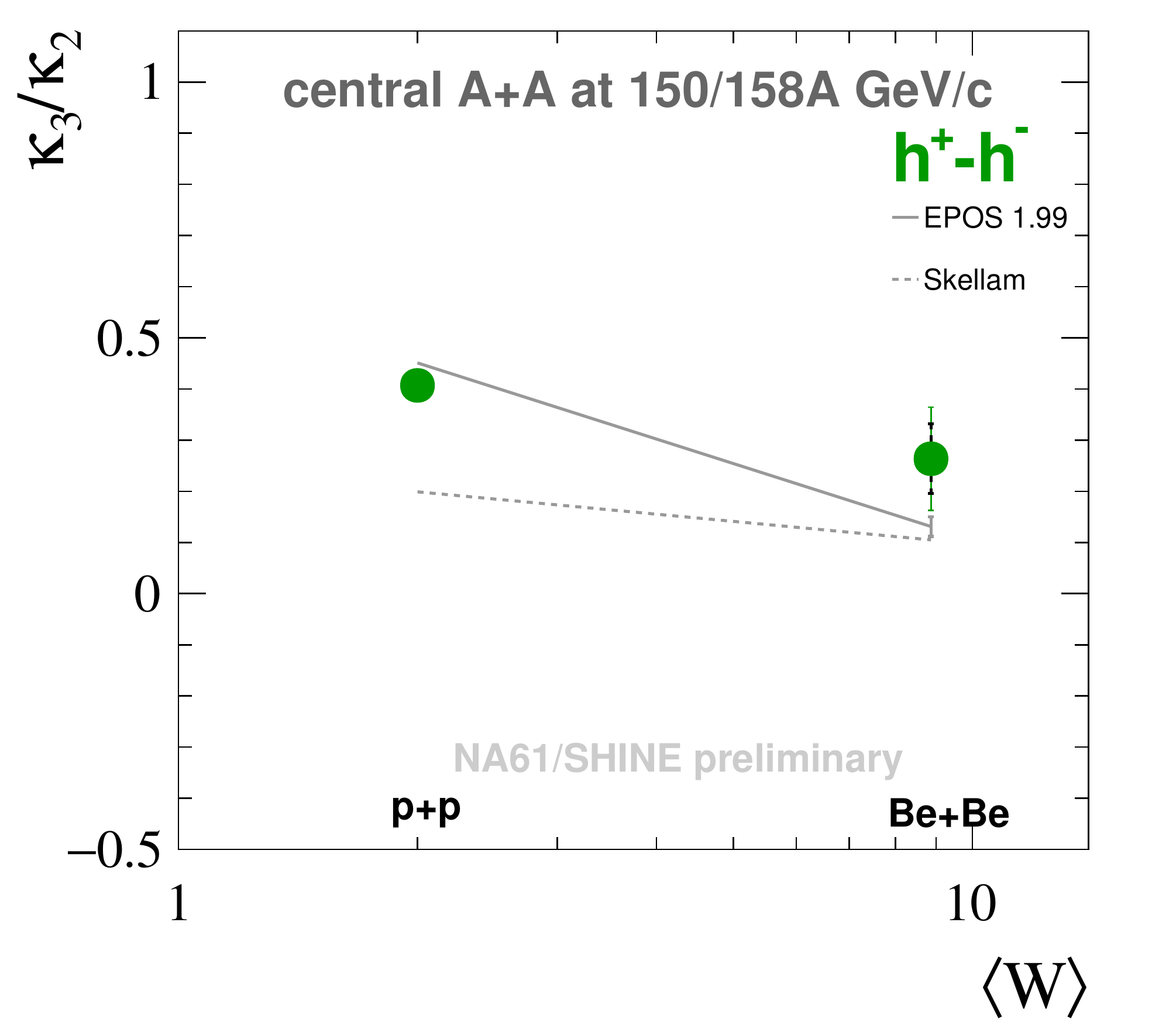}
\includegraphics[width=0.45\textwidth]{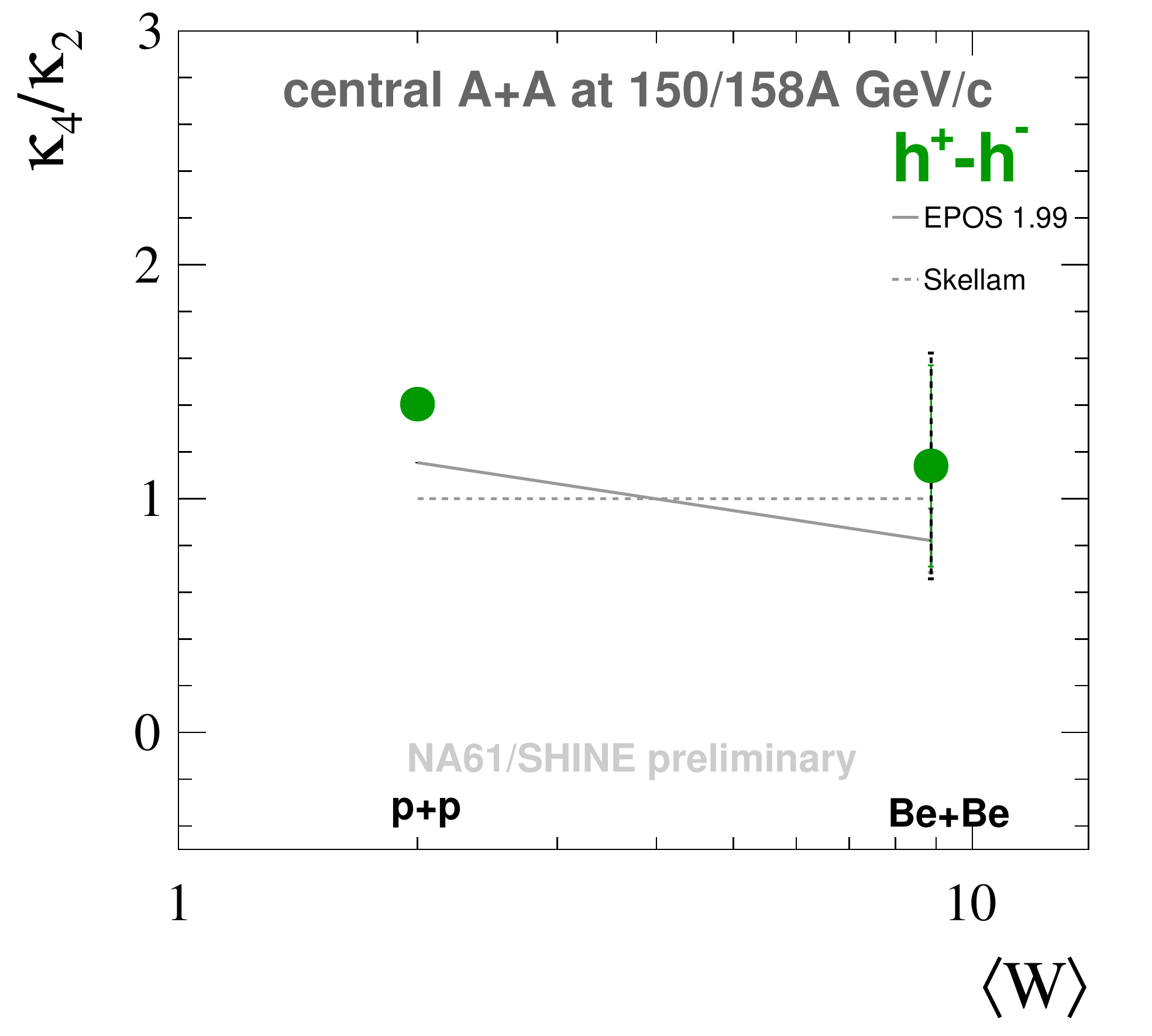}
\caption{Preliminary NA61/SHINE results on the system size dependence of $\kappa_{3}/\kappa_{2}[h^{+}-h^{-}]$ and $\kappa_{4}/\kappa_{2}[h^{+}-h^{-}]$ at 150/158$A$ GeV/$c$ as a function of the  mean number of wounded nucleons, $\langle W\rangle$.}
\label{iq}
\end{figure}
\begin{figure}[hbt]
\centering
\includegraphics[width=0.31\textwidth]{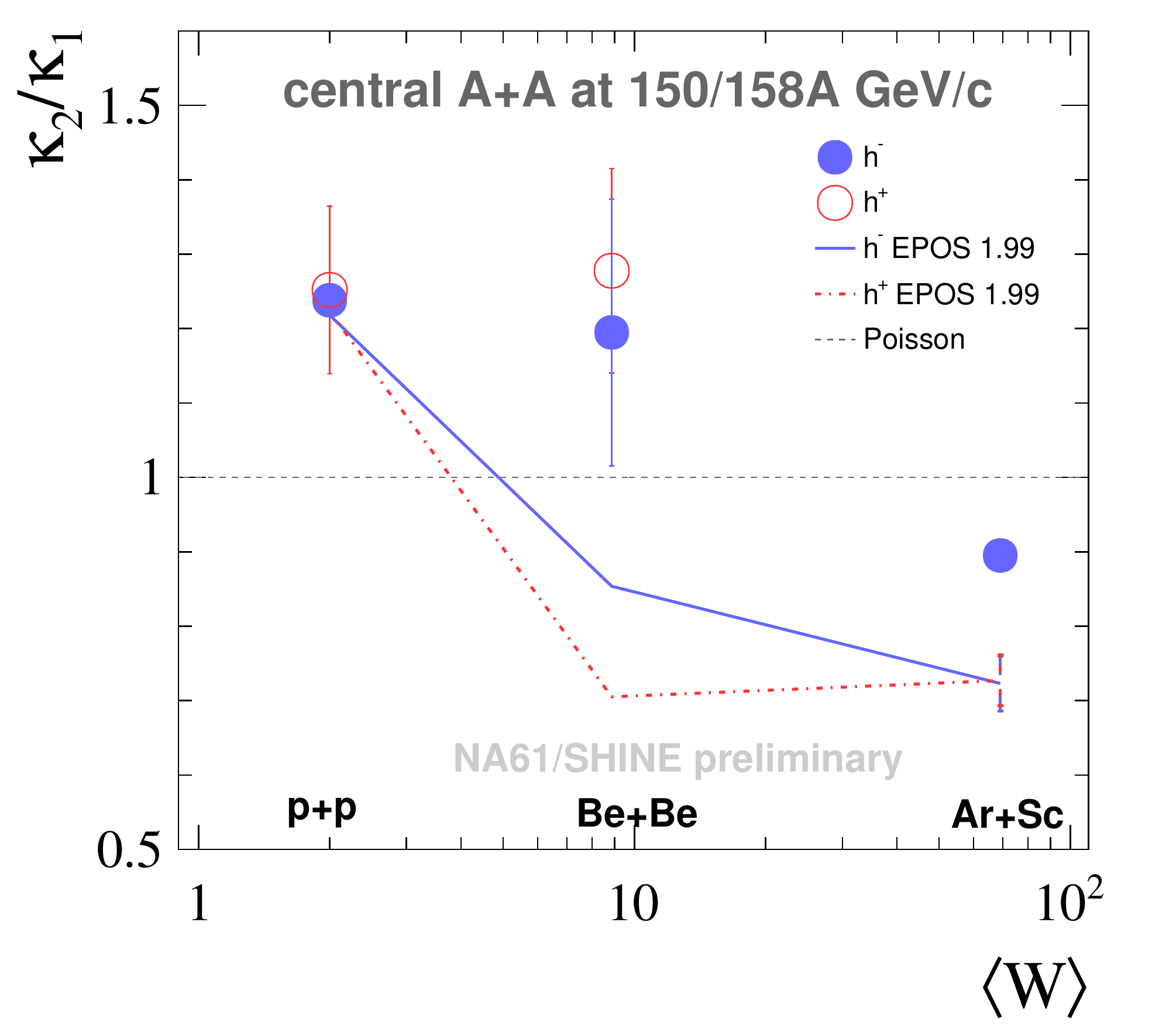}
\includegraphics[width=0.31\textwidth]{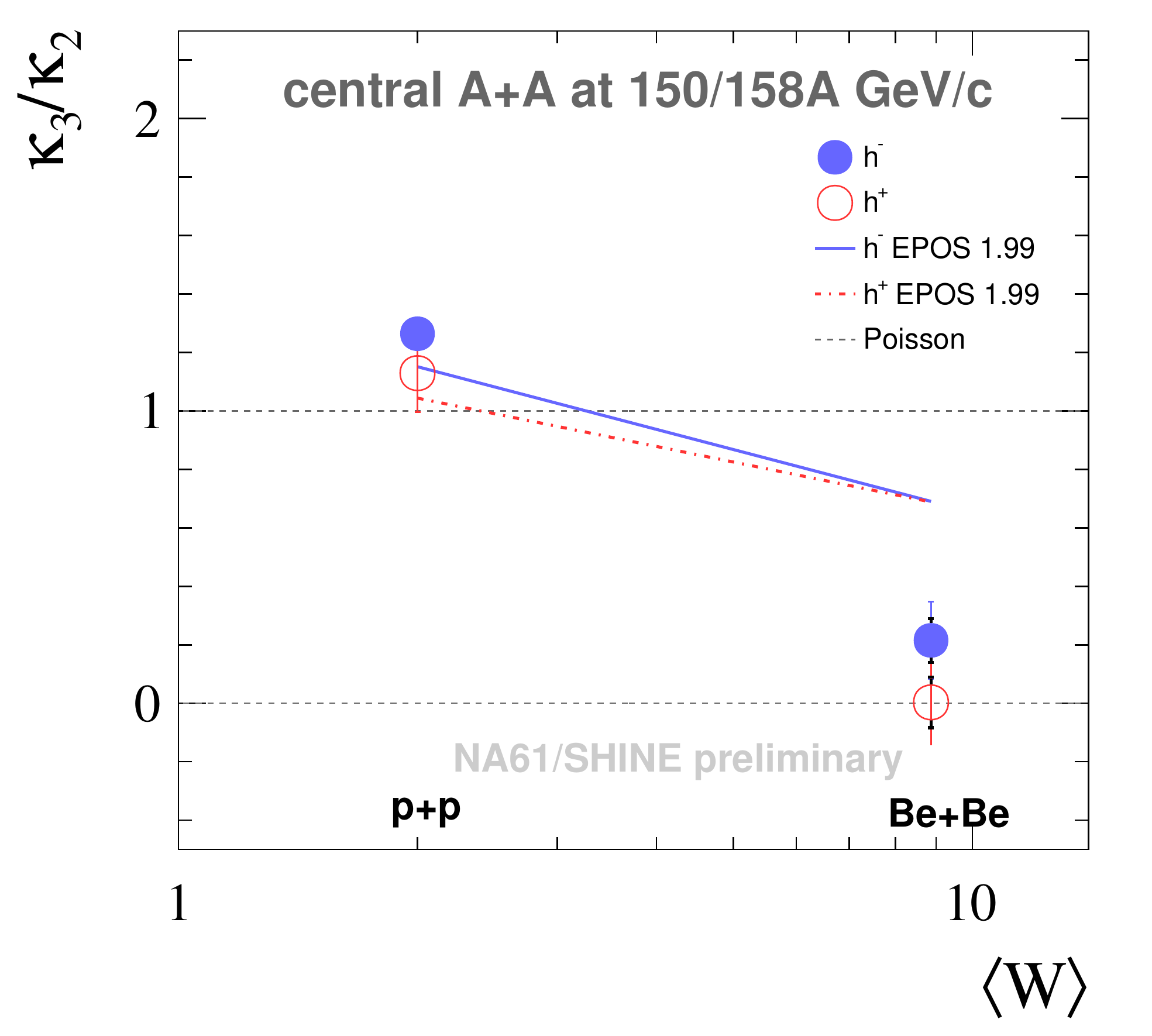}
\includegraphics[width=0.31\textwidth]{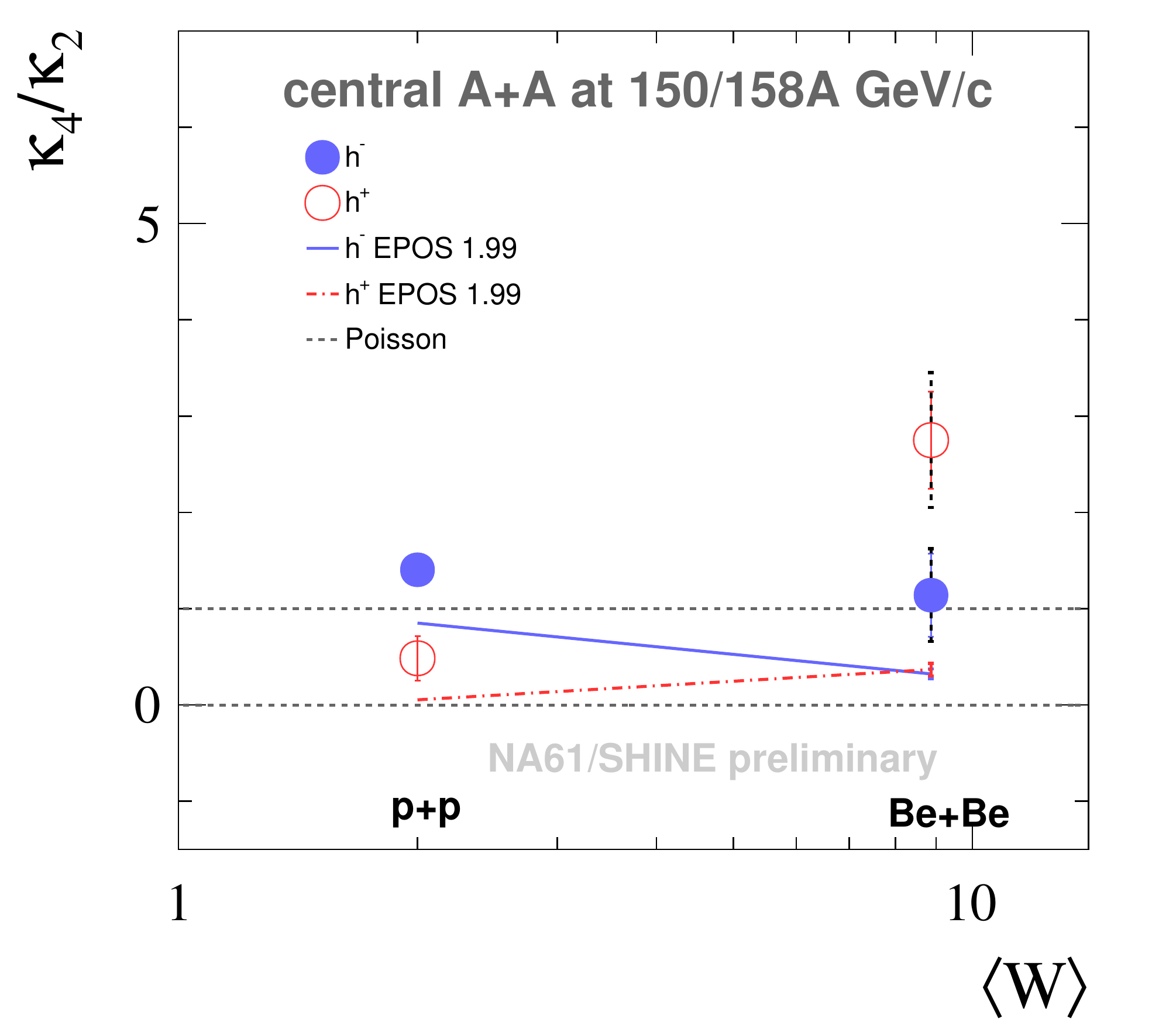}
\caption{Preliminary NA61/SHINE results on the system size dependence of multiplicity fluctuations of $h^{+}$ and $h^{-}$ at 150/158$A$ GeV/$c$ as a function of the  mean number of wounded nucleons, $\langle W\rangle$.}
\label{iqch}
\end{figure}
The expected signal of a critical point is a non-monotonic dependence of various fluctuation/correlation measures in such a scan. A specific property of the CP - the increase in the correlation length - makes fluctuations its basic signal. Special interest is devoted to fluctuations of conserved charges (electric, strangeness or baryon number)~\cite{Stephanov_overview, Asakawa:2015ybt}.

In order to compare fluctuations in systems of different sizes one should use intensive quantities, i.e. quantities insensitive to system volume. Such quantities are constructed by division of cumulants $\kappa_{i}$ of the measured distribution (up to fourth order), where $i$ is the order of the cumulant. For second, third and fourth order cumulants intensive quantities are defined as: $\kappa_{2}/\kappa_{1}$, $\kappa_{3}/\kappa_{2}$ and $\kappa_{4}/\kappa_{2}$.

Figure~\ref{iq} shows the system size dependence of third and fourth order cumulant ratio of net-electric charge at 150/158$A$ GeV/$c$. 
Measured data are in agreement with EPOS 1.99 model~\cite{Pierog:2009zt,EPOSWeb} predictions. More detailed examination of system size dependence of the same quantities for negatively and positively charged hadrons (Fig.~\ref{iqch}) shows very different system size dependence. Moreover, none of the measured quantities of $h^{+}$ and $h^{-}$ are reproduced by the EPOS 1.99 model. This disagreement indicates that we do not fully understand the underling physics how fluctuations are induced. Thus, more detailed studies are needed.

In search of CP a possible tool is proton intermittency which should follow power-law fluctuations near CP. It can be checked by studying the scaling behaviour of 2$^{nd}$ factorial moments $F_{2}(M)$ with the cell size or, equivalently, with the number of cells in ($p_x$,$p_y$) space of protons at mid-rapidity (see Refs.~\cite{Bialas:1985jb,Turko:1989dc,Diakonos:2006zz}). For experimental data a non-critical background must be subtracted with mixed events. After subtraction, the second factorial moments $\Delta F_2(M)$ should scale according to power-law for $M>>1$ with resulting critical exponent $\phi_2$ comparable to theoretical predictions~\cite{Antoniou:2006zb}. 
\begin{figure}[ht]
\centering
\includegraphics[width=0.45\textwidth]{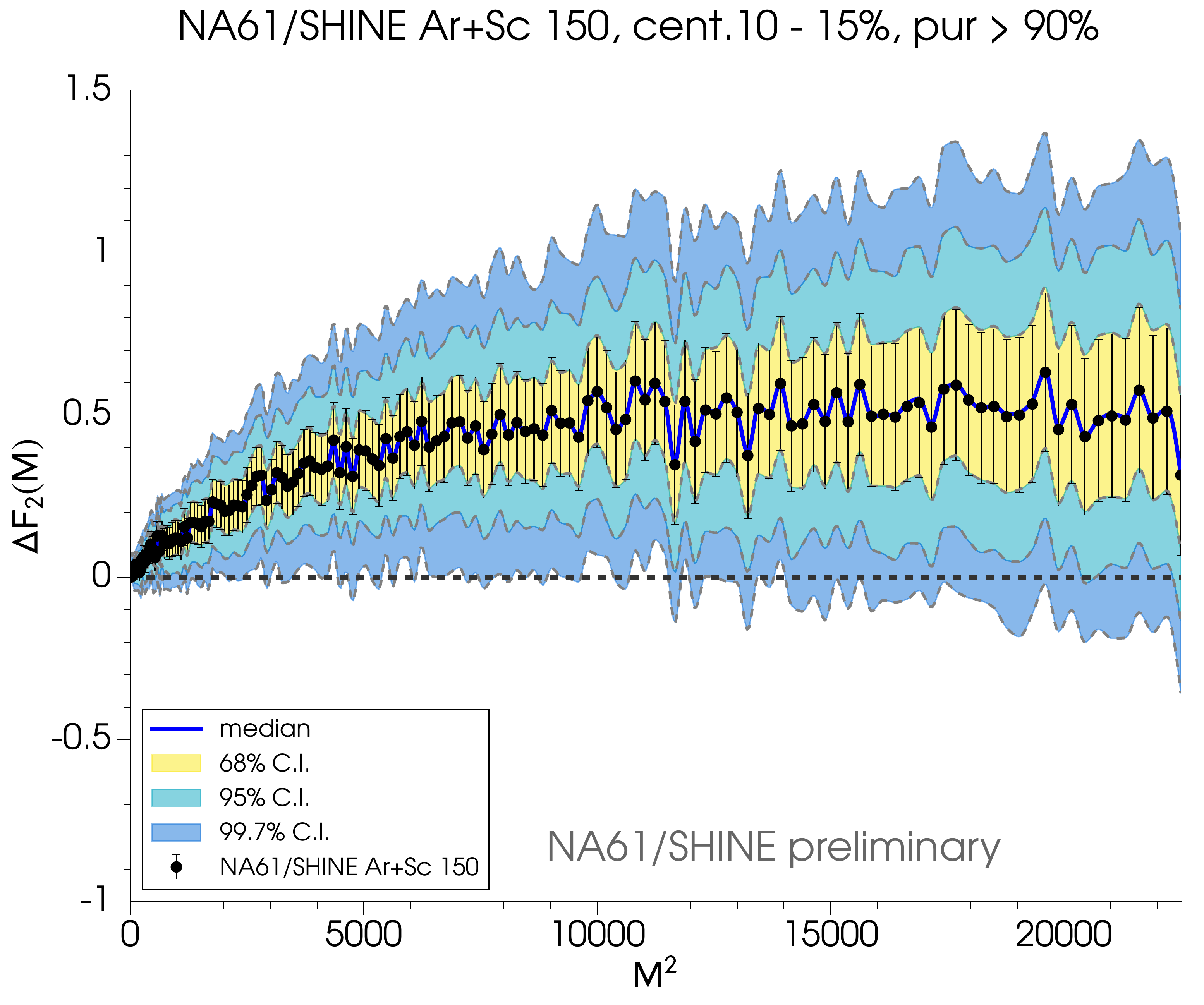}
\includegraphics[width=0.45\textwidth]{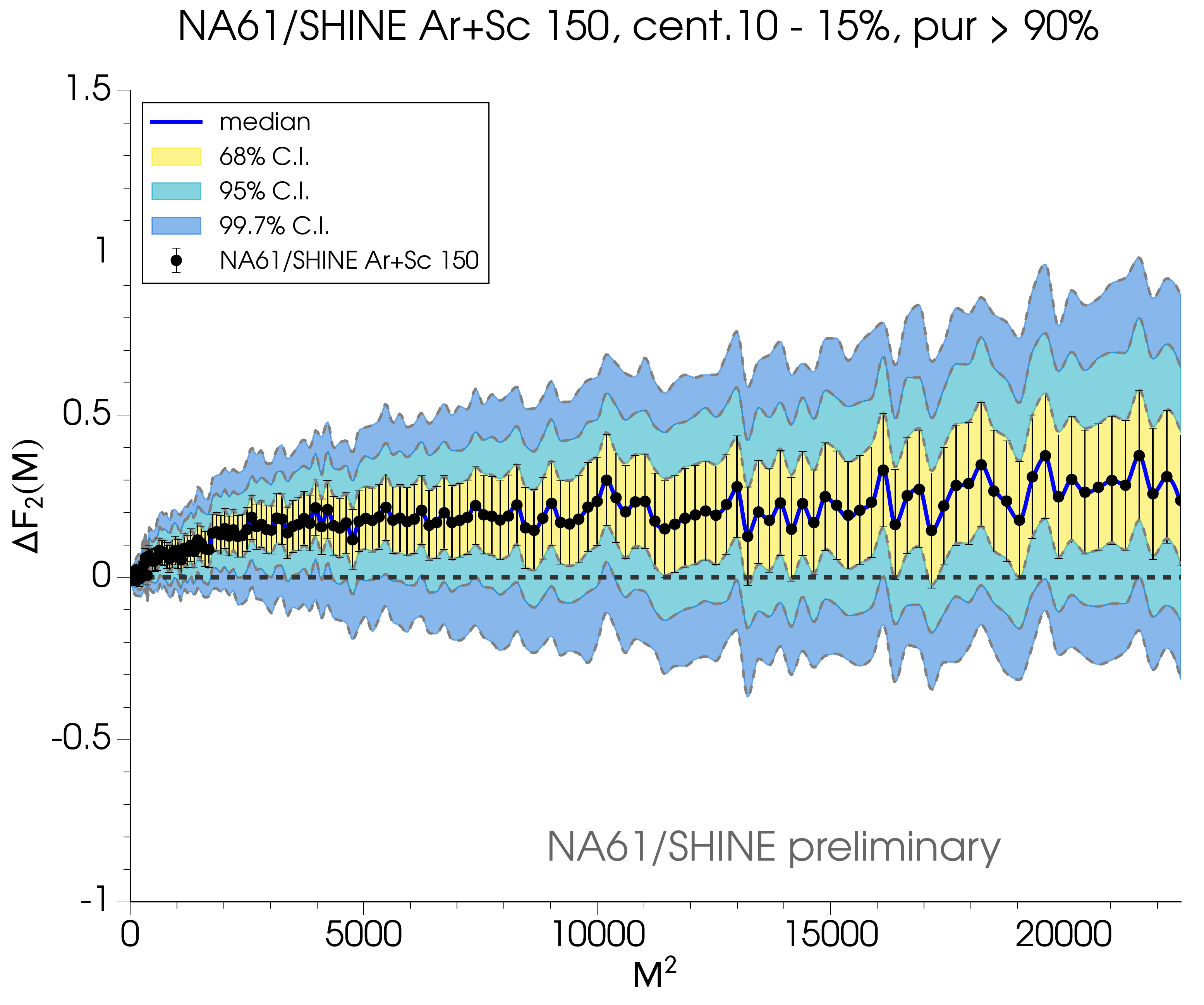}
\caption{Comparison of preliminary NA61/SHINE results on proton intermitency signal with lower (left) and higher (right) statistics}
\label{inter}
\end{figure}
Figure~\ref{inter} shows $\Delta F_2 (M)$ in semi-central Ar+Sc interactions at 150$A$ GeV/$c$. The difference between left and right side of the figure is the considered statistics. Left hand side shows results released in 2018~\cite{Davis:2019vnk}. These results indicate positive values of $\Delta F_2$ which may be connected with the CP. Right hand side shows the same results but with higher statistics (208k events vs 143k events) where $\Delta F_2$ signal is weaker.

\begin{figure}
\centering
\includegraphics[width=0.45\textwidth]{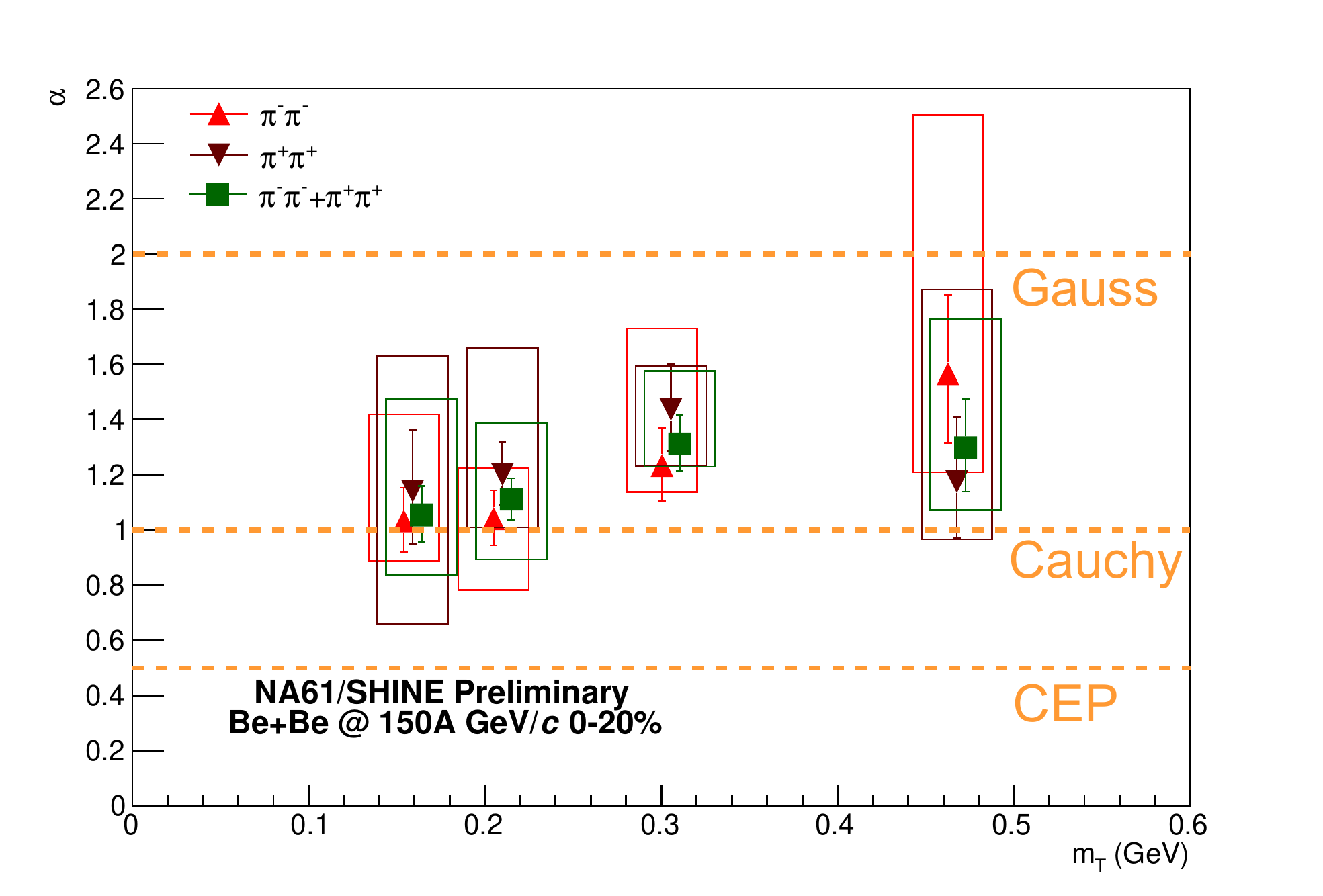}
\includegraphics[width=0.41\textwidth]{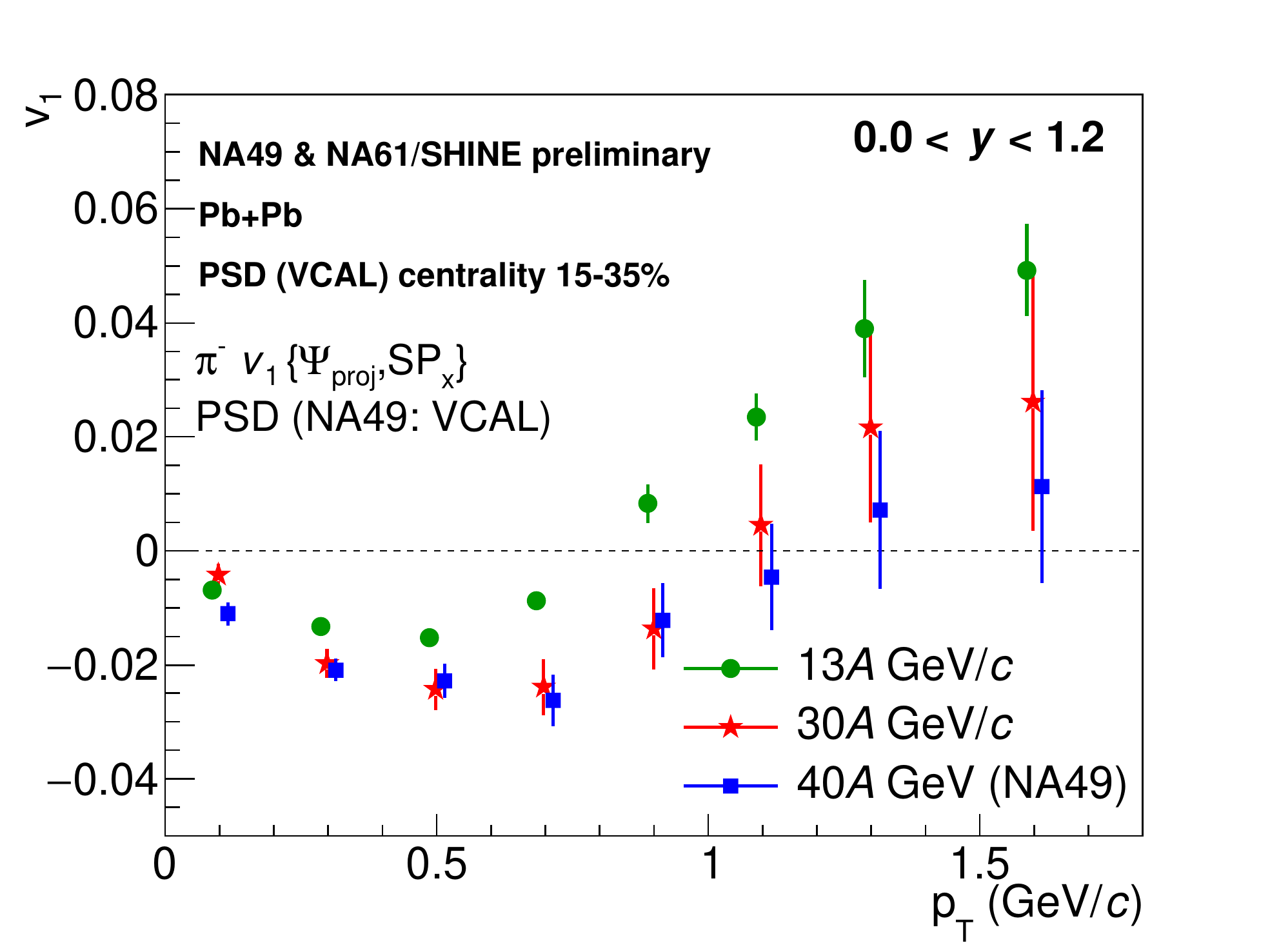}
\caption{Left: Preliminary NA61/SHINE results on $\alpha$ parameter in Be+Be collisions at 150$A$ GeV/$c$. Right: NA49~\cite{Golosov:2019sdu} and NA61/SHINE~\cite{Klochkov:2018xvw} preliminary results on energy dependence of $v_{1}$ in Pb+Pb collisions.}
\label{hbt}
\end{figure}
At the CP, the spatial correlation function becomes a power-law $\sim r^{-(d-2+\eta)}$, where $d$ represents the number of  dimensions. 
One can predict a critical exponent $\eta$ (related to spatial correlations) for the QCD universality class, which is the 3D-Ising model for QCD~\cite{Stephanov:1998dy, Halasz:1998qr}. The predicted value of $\eta$ at the CP is 0.03631~\cite{El-Showk:2014dwa}. For the random field 3D Ising $\eta=0.50\pm0.05$~\cite{Rieger:1992th}. 
In the HBT analysis the momentum correlation function $C(q)$ of produced particles is directly related to the normalized source distribution $S(r)$ 
via $C(q) = 1 + (|\tilde{S}|)^2$, where $\tilde{S}$ is the Fourier transformation of $S(r)$. 
The data analysis was done by using a Levy distributed source function~\cite{Porfy:2019dtc}.  
Since, it leads to the same power-law tails, the Levy exponent $\alpha$ was assumed to be identical to the spatial correlation exponent $\eta$~\cite{Csorgo:2003uv}.  
In the vicinity of the critical point, very low $\alpha$ values (around 0.5) 
may be expected and this can be measured by investigating the Bose-Einstein correlation function $C(q)=1+\lambda e^{-(qR)\alpha}$. 
Figure~\ref{hbt} (left) shows measured values of $\alpha$ parameter for pion pairs ($\pi^{-}\pi^{-}$, $\pi^+\pi^+$ and $\pi^{-}\pi^{-}+\pi^+\pi^+$) in 20$\%$ most central Be+Be collisions at 150$A$ GeV/$c$. 

All measured combinations indicate $1<\alpha<2$ which is far from the CP value. In addition to the CP, $\alpha$ values lower then 2 can be caused by anomalous diffusion, QCD fractal structured jet fragmentation, and also to some extend by the averaging over broad event class (e.g. centrality)~\cite{Csanad:2007fr, Csorgo:2004sr, Csorgo:2005it, Cimerman:2019hva}.

\section{Bulk matter properties}
\label{sec:OD}
Spatial asymmetry of the initial energy density in the overlaping region of the colliding relativistic nuclei is converted, via interactions between produced particles, to the asymmetry of momentum distribution of particles in the final state. The resulting asymmetry encodes important information about the transport properties of the QCD matter created during the collision. Asymmetry is usually quantified with the coefficients $v_n$ in a Fourier decomposition of the azimuthal distribution of produced particles relative to the reaction plane. The NA61/SHINE has an unique way to estimate the reaction plane with the Projectile Spectator Detector (for details see Refs.~\cite{Golosov:2019sdu,EKashirin}). 

The energy dependency of flow coefficients is of particular importance.  At the SPS energies it is expected that the slope of proton directed flow at mid-rapidity, $dv_1/dy$, changes its sign~\cite{STARv,STARv2,Wu:2018qih}. Directed flow of $\pi^{-}$ for Pb+Pb collisions at different energies is presented in Fig.~\ref{hbt} (right). The NA49 results~\cite{Golosov:2019sdu} for 40$A$ GeV/$c$ were obtained using spectator plane estimated with Veto calorimeter (VCAL). Directed flow of $\pi^{-}$, in particular the $p_T$ dependence where $v_1$ changes its sign, shows collision energy dependence. 

\section{Summary}
\label{sec:sum}

Numerous experimental results show no indications of the CP in Be+Be collisions at 150$A$ GeV/$c$. Enlarged statistics of Ar+Sc interactions has not improved the significance of the proton intermitency signal, making it inconclusive. In order to qualitatively measure the CP signal, the background phenomena should be studied in details. Measured system size dependences of multiplicity fluctuations show clear disagreement with model predictions.

The measured directed flow of pions shows energy dependence, with the slope of negatively charged pions changing sign at different collision centralities in Pb+Pb collisions (for details see Ref.~\cite{EKashirin}).
$\quad$\\

\textbf{Acknowledgements:} This work was supported by the National Science Centre, Poland under grants no. 2016/21/D/ST2/01983 and 2014/14/E/ST2/00018.











\end{document}